# The Revolution Yet to Happen

C. Gordon Bell

Jim Gray

# THE REVOLUTION YET TO HAPPEN

## Gordon Bell and Jim Gray
## Bay Area Research Center, Microsoft Corp.

## Abstract

By 2047, almost all information will be in cyberspace (1984) -- including all knowledge and creative works. All information about physical objects including humans, buildings, processes, and organizations will be online. This trend is both desirable and inevitable. Cyberspace will provide the basis for wonderful new ways to inform, entertain, and educate people. The information and the corresponding systems will streamline commerce, but will also provide new levels of personal service, health care, and automation. The most significant benefit will be a breakthrough in our ability to remotely communicate with one another using all our senses.

The ACM and the transistor were born in 1947. At that time the stored program computer was a revolutionary idea and the transistor was just a curiosity. Both ideas evolved rapidly. By the mid 1960s integrated circuits appeared -- allowing mass fabrication of transistors on silicon substrates. This allowed low-cost mass-produced computers. These technologies enabled extraordinary increases in processing speed and memory coupled with extraordinary price declines.

The only form of processing and memory more easily, cheaply, and rapidly fabricated is the human brain. Peter Cohrane (1996) estimates the brain to have a processing power of around 1000 million-million operations per second, (one Petaops) and a memory of 10 Terabytes. If current trends continue, computers could have these capabilities by 2047. Such computers could be "on body" personal assistants able to recall everything one reads, hears, and sees.

## Introduction

For five decades, progress in computer technology has driven the evolution of computers. Now they are everywhere: from mainframes to pacemakers; from the telephone network to carburetors. These technologies have enabled computers to supplement and often supplant other information processors, including humans. In 1997 processor speed, storage capacity, and transmission rate are evolving at an annual rate of 60% (doubling every 18 months, or 100 times per decade).

It is safe to predict the computers at ACM 2047 will be at least 100,000 times more powerful than those of today[1]. However, if processing, storage, and network technologies continue to *evolve* at the annual factor of 1.60 rate known as Moore's Law (Moore, 1996), then the computers at ACM 2047 will be 10 billion times more powerful than those of today!

A likely path, clearly visible in 1997, is the creation of thousands of essentially zero cost, specialized, system-on-a-chip computers we call MicroSystems. These one chip, fully-networked, systems will be everywhere embedded in everything from phones, light switches, motors, and building walls. They'll be the eyes and ears for the blind and deaf. On-board networks of them will "drive" vehicles that communicate with their counter-parts embedded in highways and other vehicles. The only limits will be our ability to interface computers with the physical world – i.e. the interface between cyberspace and physical space.

---

[1]The Semetech (1994) National Semiconductor Roadmap, predicts that by 2010 a factor of 450 more transistors will reside on a chip than in 1997. This is based on an annual growth in transistors per chip of a factor of 1.6. Only a factor 225, or an annual improvement of 1.16 would required over the remaining 37 years.

Algorithm speeds have improved at the same rate as hardware, measured in operations to carry out a given function or generate and render an artificial scene.  This double hardware-software acceleration further shortens the time it will take to reach the goal of a fully cyberized world.

This chapter's focus may appear conservative because it is based on extrapolations of clearly established trends.   It assumes no major discontinuities, and assumes more modest progress than the last 50 years.  It isn't based on quantum computing, DNA breakthroughs, or unforeseen inventions.   It does assume serendipitous advances in materials, and micro-electromechanical systems (MEMS) technology.

 Past forecasts by one of us (GB) about software milestones such as computer speech recognition tended to be optimistic.  The technologies usually took longer than expected.  On the other hand, hardware forecasts have usually been conservative.  For example, in 1975, as head of R & D at Digital Equipment, Bell forecast that a $1,000,000 eight megabyte, time-shared computer system would sell for $8,000 in 1997, and that a single user 64 kilobyte system such as an organizer or calculator would sell for $100.  While these 22 year old predictions turned out to be accurate, Bell failed to predict that high volume manufacturing would further reduce prices and enable sales of 100 million personal computers per year.

Vannevar Bush (1945) was prophetic about the construction of a hypertext based, library network.  He also outlined a speech to printing device and head mounted camera.  Charles Babbage was similarly prophetic in describing digital computers.   Both Bush and Babbage were rooted in the wrong technologies.  Babbage thought in terms of gears.   Bush's Memex based on dry photography for both storage and retrieval was completely impractical. Nonetheless, the inevitability and fulfillment of Babbage's and Bush's dreams have finally arrived. The lesson from these stories is that our vision may be clear but our grasp of future technologies is probably completely wrong.

The evolution of the computer from 1947 to the present is the basis of a model that we will use to forecast computer technology and its uses in the next five decades. We believe our quest is to get all knowledge and information into cyberspace.  Indeed, to build the ultimate computer that complements "man".

## *A View of Cyberspace*

Cyberspace will be built from three kinds of components (as diagrammed in figure 1)

- *computer platforms and the content they hold* *made of processors, memories, and basic system software;*

- *hardware and software interface transducer technology*  *that connects platforms to people and other physical systems; and*

- *networking* *technology for computers to communicate with one another.*

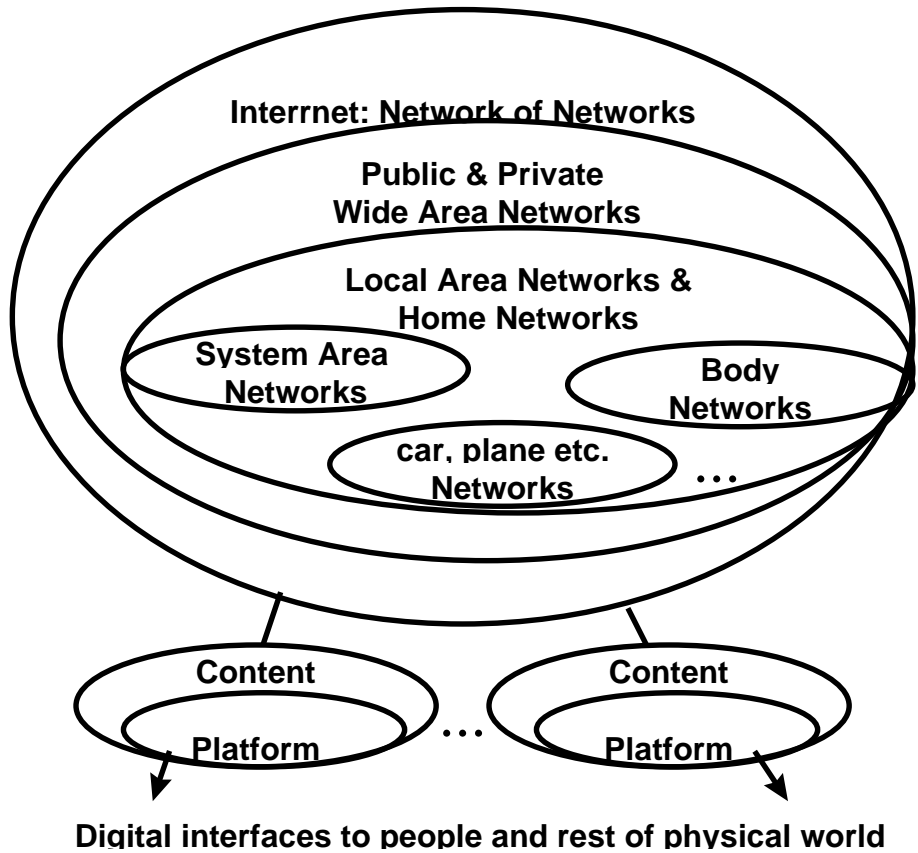

*Figure 1. Cyberspace consists of a hierarchy of networks that connects computer platforms that process, store, and interface with the cyberspace user environments in the physical world.*

The functional levels that make up the infrastructure for constructing cyberspace of Figure 1 are given in Table 1.

*Table 1. Functional Levels of the Cyberspace Infrastructure.*

| | | |
|---|---|---|
| 6 | **cyberspace user environments** | mapped by geography, interest, and demography for commerce, education, entertainment, communication, work and information gathering |
| 5 | **content** | e.g. intellectual property consisting of programs, text, databases of all types, image, audio, video, etc. that serve the corresponding user environments |
| 4 | **applications** | for human and other physical world use that enable content creation |
| 3 | **hardware & software computing platforms <u>and</u> networks** | |
| 2 | **hardware components** | e.g. microprocessors, disks, transducers interfacing to physical world, network links |
| 1 | **materials and phenomena** | (e.g. silicon) for components |

With increased processing, memory, and ability to deal with more of the physical world, computers have evolved to handle more complex data-types. The first computers only handled scalars and simple records.

With time, they have evolved to work with vectors, complex databases, graphical objects for visualization, and time varying signals used to understand speech. In the next few years, they will deal with images, video and provide virtual reality (VR)[2] for synthesis (being in artificially created environments such as an atomic structure, building, or space craft) and analysis (recognition).

All this information will be networked, indexed, and accessible by almost anyone, anywhere, at anytime -- 24 hours a day, 365 days a year. With more complex data-types, the performance and memory requirement increase as shown in Table 2. Going from text to pictures to video demands performance increases in processing, network speed and file memory capacity by a factor of 100 and 1000, respectively. Table 2 gives the memory necessary for an individual to record everything he/she read, heard, and saw during their lifetime. This varies by a factor of 40,000 from a few gigabytes to one Petabyte (PB) – a million gigabytes .

*Table 2. Data-rates and storage requirements per hour, day, and lifetime for a person to record all the text they've read, all the speech they've heard, and all the video they've seen*

| Data-type | data-rate | storage needed per hour and day | storage needed in a lifetime |
|---|---|---|---|
| **read text, few pictures** | 50 B/s | 200 KB; 2 -10 MB | 60 - 300 GB |
| **speech text @120 wpm** | 12 B/s | 43 KB; 0.5 MB | 15 GB |
| **speech compressed** | 1 KB/s | 3.6 MB; 40 MB | 1.2 TB |
| **video compressed** | 0.5 MB/s | 2 GB; 20 GB | 1 PB |

We will still live in towns, but in 2047 we will be residents of many "virtual villages and cities" in the cyberspace sprawl defined by geography, demographics, and intellectual interests.

Multiple languages are natural barrier to communication. Much of the world's population is illiterate. Video and music, including gestures, is a universal language and easily understood by all. Thus, images, music, and video coupled with computer translation of speech may become a new, universal form of communication.

Technological trends of the past decade allow us to project advances that will significantly change society. The PC has made computing affordable to much of the industrial world and is becoming accessible the rest of the world. The Internet has made networking useful and will become ubiquitous as telephones and television become "network" ready. Consumer electronics companies are making digital video authoring affordable and useful. By 2047, people will no longer be just viewers and simple communicators. Instead, we'll all be able to *create* and *manage* as well as *consume* intellectual property. We will become symbiotic with our networked computers for home, education, government, health care and work; just as the industrial revolution was symbiotic with the steam engine and later electricity and fossil fuels.

Let's examine the three cyberspace building blocks: platforms, hardware and software cyberization interfaces, and networks. Various environments such as the ubiquitous "do what I say" interface will be given and the reader is invited to create their own future scenario.

---

[2] Virtual Reality is an environment that couples to the human senses: sound, 3D video, touch, smell, taste, etc.

# Computer Platforms: The Computer and Transistor Revolution

Two forces drive the evolution in computer technology: (1) the discovery of new materials and phenomenon, and (2) advances in fabrication technology . These advances enable new architectures and new applications. Each stage touches a wider audience. Each stage raises aspirations for the next evolutionary step. Each stage stimulates the discovery of new applications that drive the next innovative cycle.

## *Hierarchies of logical and physical computers: many from one and one from many*

One essential aspect computers is that they are universal machines. Starting from a basic hardware interpreter, "virtual computers" can be built on top of a single computer in a hierarchical fashion to create more complex, higher-level computers. A system of arbitrary complexity can thus be built in a fully layered fashion. The usual levels are as follows. First a micro-machine implements an Instruction-Set Architecture (ISA). Above this is layered a software operating system to virtualize the processors and devices. Programming languages and other software tools, further raise the abstraction level. Applications like word processors, spreadsheets, database managers, and multi-media editing systems convert the systems to tools directly useable by content authors. These authors are the ones who create the real value in cyberspace: the analysis and literature, the art and music, and movies, the web sites, and the new forms of intellectual property emerging in the Internet.

It is improbable that the homely computer built as a simple processor-memory structure will change. It is most likely to continue on its evolutionary path with only slightly more parallelism, measured by the number of operations that can be carried out per instruction. It is quite clear that one major evolutionary path will be the multitude of nearly zero cost, MicroSystem (system-on-a-chip) computers customized to particular applications.

Since one computer can simulate one or more computers, multiprogramming is possible where one computer provides many computers to be used by one or more persons (timesharing) doing one or more independent things via independent processes. Timesharing many users on one computer was important when computers were very expensive. Today, people only share a computer if that computer has some information that all the users want to see.

The multi-computer is the opposite of a time-shared machine. Rather than many people per computer, a multi-computer has many computers per user. Physical computers can be combined to behave as a single system far more powerful than any single computer.

Two forces drive us to build multi-computers. (1) Processing and storage demands for database servers, web servers, and virtual reality systems exceed the capacity of a single computer. At the same time, (2) the price of individual computers has declined to the point that even a modest budget can afford to purchase a dozen computers. These computers may be networked to form a distributed system. Distributed operating systems using high-performance low-latency System Area Networks (SANs) can transform a collection of independent computers into scalable *cluster* that can perform large, computational and information serving tasks. These clusters can use the spare processing and storage capacity of the nodes to provide a degree of fault-tolerance. Clusters become the server nodes of the distributed, worldwide Intranets. All Intranets tie together forming the Internet.

The commodity computer nodes will be the cluster building blocks – we call them *CyberBricks* (Gray, 1996). By 2010, Sematech predicts CyberBricks with memories of 30 gigabytes, made from 8 gigabyte memory chips and processing speeds of 15 giga instructions per second (Semetech, 1994).

Massive computing power will come via scalable clusters of CyberBricks In 1997, the largest, scalable clusters contain hundreds of computers. Such clusters are used for both commercial database and transaction processing and for scientific computation. Meanwhile, large scale multiprocessors that maintain a coherent shared memory seem limited to a few tens of processors, and have very high unit-costs. For 40 years, researchers have attempted to build scalable, shared memory multiprocessors with over 50 processors, but this goal is still elusive. Certainly they have built such machines, but the price and performance have been disappointing. Given the low cost of single chip or single substrate computers – it appears that large-scale multi-processors will find it difficult to compete with clusters built from CyberBricks.

## *Semiconductors: Computers in all shapes and sizes*

While many developments have permitted the computer to evolve rapidly, the most important gains have been semiconductor circuit density increases and storage density in magnetics measured in bits stored per square inch. In 1997, these technologies provide an annual 1.6 fold increase. Due to fixed costs in packaging and distribution, prices of fully configured systems improve more slowly, typically 20% per year. At this rate, the cost of computers commonly used today will be $1/10^{th}$ of their current prices in 10 years.

Density increases enable chips to operate faster and cost less, because:

- The smaller everything gets, approaching the size of an electron, the faster the system behaves.

- Miniaturized circuits produced in a batch process tend to cost very little once the factory is in place. The price of a semiconductor factory appears to double with each generation (3 years). Still, the cost per transistor declines with new generations because volumes are so enormous.

Figure 2 shows how the various processing and memory technologies could evolve for the next 50 years. The Semiconductor industry makes the analogy that if cars evolved at the rate of semiconductors, today we would all be driving Rolls Royces that go a million miles an hour and cost $0.25. The difference here is that computing technology operates Maxwell's equations defining electromagnetic systems, while most of the physical world operates under Newton's laws defining the movement of objects with mass.

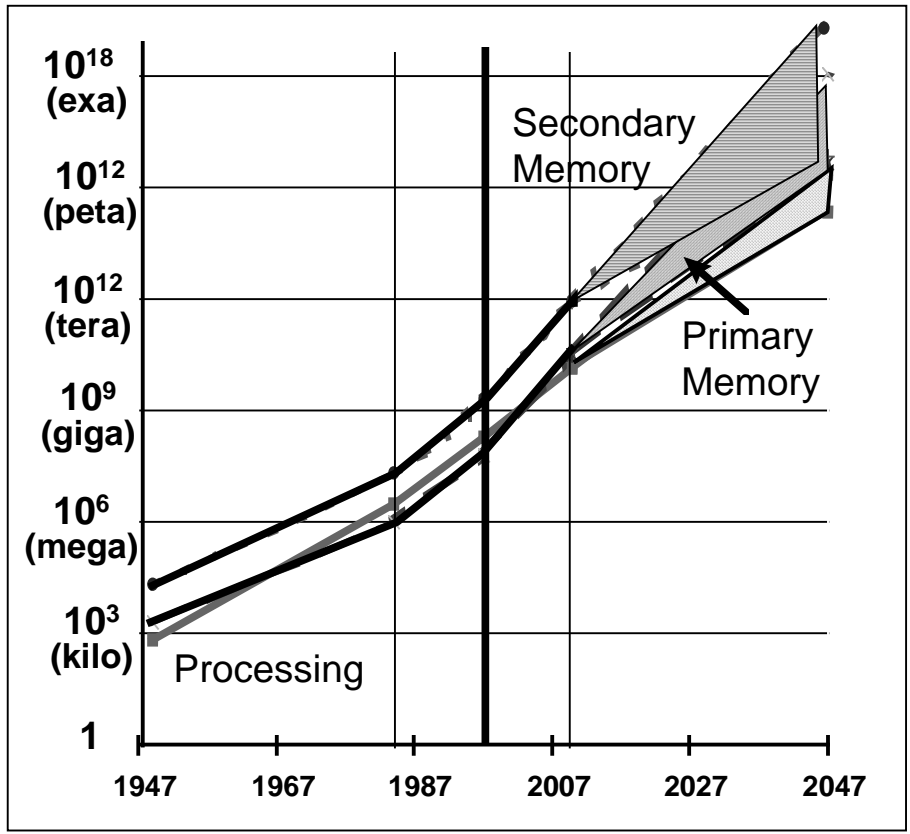

*Figure 2. Evolution of computer processing speed in instructions per second, and primary and secondary memory size in bytes from 1947 to the present, with a surprise-free projection to 2047. Each division is three orders of magnitude and occurs in roughly 15-year steps.*

In 1958, when the integrated circuit (IC) was invented, until about 1972, the number of transistors per chip doubled each year. In 1972, the number began doubling only every year and a half, or increasing at 60 percent per year, resulting in a factor of 100 improvement each decade. Consequently, each three years semiconductor memory capacities have increased four fold. This phenomenon is known as Moore's Law, after Intel's Founder and Chairman, Gordon Moore, who first observed and posited it.

Moore's Law is nicely illustrated by the number of bits per chip of dynamic random-access memory (DRAM) and the year in which each chip was first introduced: 1K (1972), 4K (1975), 16K (1978), … 64M (1996). This trend is likely to continue until 2010. The National Semiconductor Roadmap (Semetech, 1994) calls for 256 Mbits or 32 Mbytes next year, 128 Mbytes in 2001, … and 8 GBytes in 2010!

## *The Memory Hierarchy*

Semiconductor memories are an essential part of the memory hierarchy because they match processor speeds. A processor's small, fast registers hold a program's current data and operate at processor speeds. A processor's, larger, slower cache memory built from static RAM (SRAM) holds recently used program and data that come from the large, slow primary memory DRAMs. Magnetic disks with millisecond access times form the secondary memory that holds files and databases. Electro-optical disks and magnetic tape with second and minute access times are used for backup and archives that form the tertiary memory. This memory hierarchy operates because of temporal and spatial locality, whereby recently used information is likely to be accessed in the near future, and a block or record that is brought into primary memory from secondary memory is likely to have additional information that will be accessed.

Note that each lower level in this technological hierarchy is characterized by slower access times, and more than an order of magnitude lower cost per bit stored.   It is essential that each given memory type improve over time, or else it will be eliminated from the hierarchy.

Just as increasing transistor density has improved the storage capacity of semiconductor memory chips, increasing areal density[3] has directly affected the total information-storage capacity of disk systems. IBM's 1957 disk file, the RAMAC 350, recorded about 100 bits along the circumference of each track and each track was separated by 0.1 inch, giving an areal density of 1,000 bits per square inch.  In early 1990, IBM announced that one of its laboratories had stored 1 billion bits in 1 square inch and shipped a product with this capacity in 1996.  This technology progression of six orders of magnitude in thirty-three years amounts to a density increase at a rate of over 50 percent per year.

Increases in storage density have led to magnetic storage systems that are not only cheaper to purchase but also cheaper to own, primarily because the density increases have markedly reduced physical volume.  5 1/4 -  and 3 1/2 -inch drives can be mounted within a workstation.  These smaller disks store much more, cost much less, are much faster and more reliable, and use much less power than their ancestors.  Without such high-density disks, the workstation environment would be impossible.

In 1992, electro-optical disk technologies provide a gigabyte of disk memory at the cost of a compact audio disk, making it economically feasible for PC or workstation users to have roughly four hundred thousand pages of pure text or ten thousand pages of pure image data instantly available.  Similarly, advances in video compression using 100s of millions of operations per second permit VHS quality video to be stored on a CD.  By 2000, one CD will hold 20 Gbytes, and by 2047 we might expect this to grow to 20 Tbytes.

## Connecting to the Physical World

Basic hardware and generic transducer software technology, coupled with networking, governs the new kinds of computers and their applications as shown in Table 3.   Paper will be described as a special case because of its tremendous versatility for memory, processing, human interface, and networking.  Paper is also civilization's first computer.

The big transitions will come with the change in user interface from Windows, Icons, Mouse, and Pull-down menus (WIMP) to speech.  Directly after speech, camera input of gestures or eye movement could enhance the user interface. In the long term, visual and spatial image input from sonar, radar, and Global Position Sensing (GPS) with a world-wide exact time base coupled with radio data links will open up new portability and mobility applications.  These include robots, vehicles, autonomous appliances, and applications where the exact location of objects is required.

Speech synthesis was first used for reading to the blind and telephone response in the mid 1970s. Now speech understanding  systems are used for limited domains such as medical report generation, and everyone foresees a useful, speech typewriter by the end of the century.  Furthermore, many predict automatic natural language translation systems that take speech input in one language and translate it to speech or text in another language by 2010.

The use of the many forms of video is likely to parallel  speech, going from graphics and the synthesis of virtual scenes and sets for desktop video productions taking place at synthesized location to analysis of spaces and objects in dynamic scenes.  Computers that can "see" and operate in real time will enable surveillance with personal identification, identification of physical objects in space for mapping and virtual reality, robot and other vehicle navigation, and artificial vision.

---

[3]The amount of information that can be stored per unit area.

Table 3 gives some applications enabled by new interface transducers.

*Table 3. Interface Technologies and their Applications*

| Interface (Transducer) | Application |
|---|---|
| large, high-quality portable displays | book, catalog, directory, newspaper, report substitution and the elimination of most common uses of paper; portability, permanency, and very low power are required for massive change! |
| personal ID | security |
| speech | input to telephones, PC, network computer, telecomputer (telephone plus computer), and tv computer; useful personal organizers and assistants; appliance and home control, including lighting, heating, security; personal companions that converse and attend to various needs; |
| synthetic video | presentations and entertainment with completely arbitrary synthesized scenes, including "computed people" |
| Global Position Sensing (GPS); exact time base | "where are you, where am I?" devices; dead reckoning navigation; monitoring lost persons & things; exact time base for trading and time stamps |
| biomedical sensor/effectors | monitoring and attendance using Body Nets, artificial cochlea and retina, etc. and implanted PDAs |
| images, radar, sonar, laser ranging | room and area monitoring; gesture for control; mobile robots and autonomous vehicles; shopping and delivery; assembly; taking care of xx; artificial vision; |

## *Paper, the first stored program computer… where does it go?*

Having all IP in cyberspace implies the potential for the elimination of paper for storing and transmitting money, stock, or legal contracts, as well as books, catalogs, newspapers, music manuscripts, and reports. Paper's staying power is impressive even though it is uneconomical compared with magnetics, but within 50 years, the cost, density, and inability to search its contents or to present multimedia will force paper's demise where storage, processing, or transmission is required. High resolution, high contrast, rugged, low-cost, portable, variable sized displays have the potential to supplant some use of paper, just as email is replacing letters, memos, reports and voice messaging in many environments. With very low cost "electronic" paper and radio or infrared networks, books for example will be able to speak to us and to one another. This is nearly what the world-wide web offers today with hypertext linked documents with spoken output. However, paper is likely to be with us forever for "screen dumps" giving portability and a lasting, irreplaceable graphical user interface (GUI). We know of no technology to attack paper's broad use in 1997!

One can argue that paper and the notion of the human interpretation of paper stored programs such as algorithms, contracts (laws and wills), directions, handbooks, maps, recipes, and stories was our first computer. Paper and its human processors perform the functions of a modern computer, including processing, memory storage hierarchy from temporary to archival, means of transmission including switching via the world-wide physical distribution network, and human interface. Programs and their human interpreters are like the "Harvard" computer architecture that clearly separated program and data.

In 1997, magnetic tape has a projected lifetime of 15 years; CDs are estimated to last 50 years provided one can find the reader, Microfilm: 200 years (unfortunately, computers can't read it yet), and acid-free paper: over 500 years.

The potential to reduce the use of paper introduces a significant problem:

> How are we going to ensure accessibility of the information, including the platforms and programs we create in 50 or 500 years that our ancestors had the luck or good fortune of providing with paper? How are we even going to assure accessibility of today's HTML references over the next 5 decades?

## Networks: A convergence and interoperability among all nets

Metcalfe's Law states the total value of a network is equal to the square of the number of subscribers, while the value to a subscriber is equal to the number of subscribers. The law describes why it is essential that everyone have access to a single network instead of being subscribers on isolated networks.

Many network types are needed to fulfill the dream of cyberspace and the information superhighway. Several important ones are listed in Table 4. Figure 2 shows the change in bandwidth of two important communication links that are the basis of Wide Area Networks and the connection to them. Local Area Network bandwidth has doubled every 3 years, or increased by a factor of 10 each decade. Ethernet was introduced in the early 1980s and operated at 10 megabits per second. It was increased to 100 Mbps in 1994 and further increased to 1 Gbps in 1997.

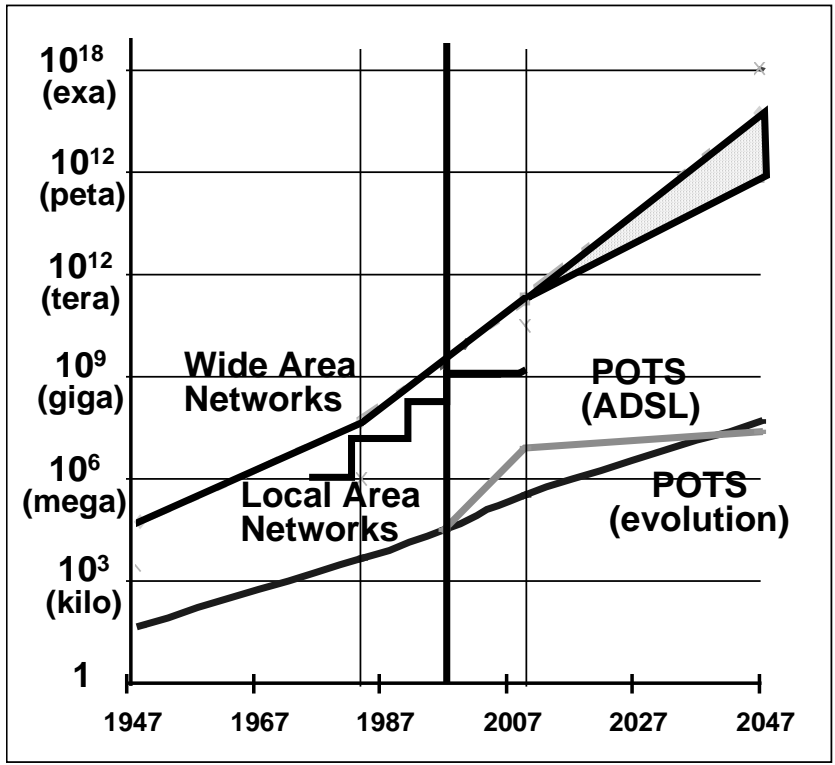

*Figure 3. Evolution of Wide Area Network, Local Area Network, and Plain Old Telephone Service (POTS) bandwidths in bits per second from 1947 to the present, and a projection to 2047.*

Four networks are necessary to fulfill the dream of cyberspace whereby all information services are provided with a single, ubiquitous, digital dial tone:

- long haul WANs that connect thousands of central switching offices,

- local loops connecting central offices to user sites via plain old telephone services (POTS) copper wires, and

- LANs and Home Networks to connect platform equipment within a site.

- wireless networks for portability and mobility

The bottleneck is the local loop or last mile (actually up to 4 miles). Certainly within five years, the solution to this problem will be clear, ranging from new fiber and wireless transmission to the use of existing cable TV and POTS. In the short term (10-25 years) installed copper wire pairs can eventually carry data at 5-20 Mbps that can encode high-resolution video. Telephone carriers are trying various digital subscriber loop technologies to relieve the bottleneck. Cable TV that uses each 6 MHz TV channel to carry up to 30 megabits per second is also being tested and deployed . Both are directed at being information providers. By 2047, fiber that carries several gigabits per optical wave length will most likely come to most homes to deliver arbitrarily high bandwidths. One cannot begin to imagine applications to utilize such bandwidth.

Once the home is reached, home networks are needed that are virtually identical to commercial LANs, but easier and cheaper to install and maintain. Within a home, the ideal solution is for existing telephony wiring to carry voice, video, and data. Telephony wiring can carry several megabits per second, but is unlikely to be able to carry the high bandwidths that high definition tv needs.

LAN and Long Haul networks are deregulated, local loop are monopolistic and regulated. By 2047 deregulation will be complete and the local loop will catch up with its two radical LAN and WAN siblings.

The short-term prospects of "one dial tone" that can access arbitrary subscribers or data sources for voice, video, and data before 2010 are not bright (Bell and Gemmell, 1996). Telephony's voice channels carry at most 64 Kbps and television is evolving to require 5 Mbps, or a factor of 100 difference. Similarly, data files are growing from the several hundred Kbytes for an hour or so of text material that one might read, to 10s of Mbytes of text with pictures, to 2 Gbytes for an hour of high quality video.

By 2047 we would hope for a "one dial tone" or single service whereby the bits are fungible and can be used for telephony, videotelephony, television, web access, security, home and energy management, and other digital services. Or will there be 2 or 3 separate networks that we have today for telephony, television, data, and other services?

*Table 4. Networks and their Application*

| Network | Technology | Application |
|---|---|---|
| Last mile (home-to-central office) | CATV, POTS lines, long-term = fiber | carry "one dial-tone" to offices and homes for telephone, videophone, tv, web access, monitoring & control of physical plant, telework, telemedicine, tele-education |
| LAN: Local Area Network | wired | connect platforms within a building |
| wLAN: wireless Local Area Network | radio & infrared confined to small areas | portable PC, PDA, phone, videophone, ubiquitous office and home accessories, appliances, health care monitors, gateway to BAN; |
| HAN: Home Net (within homes) | wire, infrared, radio | functionally identical to a LAN |
| System x Network | wired | interconnection of the platforms of system x, such as a airplane, appliance, car, copy or production machine, or robot. SANs & BANs are system networks |
| SAN: System Area Network | standard, fast, low latency | building scalables using commodity PCs and "standard" networks that can scale in size, performance, reliability, and space (rooms, campus, …wide-areas) |
| BAN: Body Net | radio | human on-body net for computation, communication, monitoring, navigation |

Wireless technology offers the potential to completely change the communications infrastructure. Therefore, a significant policy question arises around how wireless bandwidth will be allocated and potentially *reallocated* in the future. Wireless networking would allow many applications including truly portable and mobile computing, use within buildings for local and home networks, robotics, and when used with GPS, to identify the location of a platform.

Various authors have proposed a reallocation of the radio spectrum so that fixed devices such as television sets would be wired so that telephony, videotelephony, and data platforms could be mobile.

Existing radio frequency bands capable of carrying 5+ bits per hertz, could provide capacities of: 0.5Gbps (806-832 Mhz); 2.5 Gbps (<5Ghz); 1.8 Gbps (5150-5875 Mhz); and 50 Gbps (27.5-64 GHz). The actual capacity depends on the geographical cell size that enables space-sharing of a given channel that is interconnected via terrestrial cabling. Cell size depends on power, terrain, weather, and antennae location (including satellites).  For example, the Personal Handiphone System (PHS) recently deployed in Japan can communicate at a radius of 100-300 meters with each cell carrying 268 voice channels in the allocated 1895 to 1918.1 MHz band.  Digital encoding would switch about one-thousand, 8 Kbps connections – enough for Dick Tracy's low resolution wrist videophone.

The following section describes potential new platforms using the computers, interface, and network technology described above.

## Future platforms, their interfaces, and supporting networks

A theory of computer class formation posited by Bell in 1975,  (Bell and McNamara, 1991), based on Moore's Law, states that computer families follow one of three distinct  paths over time:

1. **evolution** of a class along a constant, or slightly lower price and increasing performance (and functionality) timeline.  This path is the result of a fixed cost infrastructure of suppliers and customers who benefit by having increased performance or other capabilities to track growth needs. More power allows computers to address and prototype new applications.

2. **establishment of new lower priced classes** when cost can be reduced by a factor of 10. Since price for a given function declines by about 20% per year, a new class forms about every 10 years.  The class is characterized by new hardware and software suppliers and a new style of use or new applications  for existing and new users.

3. **function commoditization as MicroSystems into appliances and other devices** whereby a function such as speech recognition, filing, printing, display are incorporated into other devices such as watches,  talking and listening calculators and phones, cameras with special graphical effects creation, pictures that interact visually and tell stories.

This theory accounted for the emergence of minicomputers (1970s) costing one hundred thousand dollars or significantly less than the original million dollar mainframes introduced in 1951, 20 thousand dollar workstations and two thousand dollar personal computers (1980s), several hundred dollar personal organizers, and ten to one hundred dollar pocket telephone book -dialers and book substitute devices such as electronic dictionaries. It also accounts for the emergence of embedded and low cost game computers using world-wide consumer content and distribution networks.

Most of us associated with computing *or any technology* use *revolution* [4] to describe something new such as the microprocessor, the PC, or personal digital assistant (PDA) because they represent a discontinuity. Since the invention of the Integrated Circuit (IC) thirty years ago, progress in these technologies has been evolutionary albeit so rapid as to look like constant revolutions. These computers are all of the same species.  They  are all based on the basic circuit and memory technologies that process and store information. Developments of new sensors and effectors (i.e. transducers) that interface to other real world systems determine how useful computers can be to process, control, store, and switch information.  And finally, in  the generation we are entering, global networking determines class formation. *Without all three components* (lower cost computer platforms, interfaces to the physical world and  users, and networks) today's computer would be just  scaled down, stand-alone mainframes that consumed tiny cards and produced much paper.

---

[4] A revolution should be a significant  "leap" that produces an even more significant benefit.

New classes have formed every 10-15 years! Table 5 gives past computer classes and those that are likely to form based on platforms, interfaces, and networks. EDSAC (1949), the first useful computer had just paper tape and a slow printer. UNIVAC (1951), the first commercial computer was fed with cards and used magnetic tape and drums for storage. IBM evolved mainframes with the System /360 (1964) to be controlled with a batch operating system and eventually to be timeshared. Timeshared computers were controlled with keyboards of alphanumeric displays. The first minicomputers (1965) were built to be embedded into other systems for control, switching, or some other function before evolving to a downsized department "mainframe". The first personal computers (1977-1981) were controlled by single user operating systems and command languages. PCs and workstations evolved to the WIMP interface previously described. More importantly, workstations required Local Area Network for inter-communication and file sharing that was inherent in a single, large timeshared computer. The first world-wide web terminals were just PCs running browser software (1993), that access a global network. In 1997 various types of low cost web access terminals, including hybrid television and telephone based terminals have been introduced using the world-wide web client-server architecture.

*Table 5.  New computer classes and the enabling components.*

| Generation | Platform (logic, memories, O/S) | User Interface and control | Network infrastructure |
|---|---|---|---|
| **The beginning (direct & batch use) (1951)** | the computer, vacuum tube, **transistor**, core, drum & mag tape | card, paper tape direct control evolving to batch O/S | none originally.. computer was stand-alone |
| **Interactive timesharing via commands; minicomputers (1965)** | <u>**integrated circuit (IC), disk**</u>, minicomputer; <u>**multiprogramming**</u> | **glass teletype** & glass keypunch, command language control | POTS using **modem,** and proprietary nets using WAN |
| **Distributed PCs and workstations (1981)** | <u>**microprocessor**</u> PCs & workstations, floppy, small disk, dist'd O/S | **WIMP (windows, icons, mouse, pull-down menus)** | **WAN, LAN** |
| **World-Wide Web access via PCs and Workstations (1994)** | Evolutionary PCs and workstations, servers everywhere, Web O/S | Browser | **fiber optics backbone, www, http** |
| **Web Computers: Network-, Tele-, TV-computers (1998)** | client software from server using JAVA, Active X, etc. | telephone, simple videophone, television access to the web | **xSDL** for POTS or cable access for hi speed data; 3 separate networks |
| **SNAP: Scalable Network & Platforms (1998)** | PC uni- or multi-processor commodity platform | server provisioning | **SAN (System Area Network) for clusters** |
| **One Info Dial tone: phone, videophone, tv, & data (2010)** | Video capable devices of all types; | video as a primary data-type | Single high speed network access; Home Net |
| **Do what I say (2001) speech controlled computers** | embedded in PCs, hand held devices, phone, PDA, other objects | **speech** | IR and radio LANs for Network access |
| **Embedding of speech & vision functions (2020)** | $1-10 of <u>chip area</u> for: books, pictures, papers, that identify themselves | | **Body Net**, Home Net, other nets |
| **Anticipatory by "observing" user behavior (2020)** | room monitoring, gesture | vision, gesture control | Home Net |
| **Body Net: vision, hearing, monitoring, control, comm., location (2025)** | artificial retina, cochlea, glasses for display, | implanted sensors and effectors for virtually every part of a body | Body Network, gateway to local IR or radio nets everywhere |
| **Robots for home, office, and factory** | general purpose robot; appliances become robot | radar, sonar, vision, mobility, arms, hands | IR and radio LAN for Home and Local Areas |

## *MicroSystems: Systems-on-a-chip*

The inevitability of complete, computer systems-on-a-chip will create a new ***MicroSystems*** industry[5]. By 2002 we expect a PC-on-a-chip with at least 32 Mbytes, video and audio I/O, built-in speech recognition, and industry standard busses for mass storage, LAN, and communication. Technology will stimulate a new industry for building applications specific computers that require partnerships among, system customers, chip fabricators, ECAD suppliers, intellectual property (IP) owners, and systems builders.

The volume of this new MicroSystem industry will be huge -- at least two orders of magnitude more units than the PC industry. For every PC, there will be thousands of other kinds of systems built around a single chip computer architecture, with its interconnection bus on chip, and that is complete with processor, memory hierarchy, i/o (including speech), firmware, and platform software. With more powerful processors, firmware will replace hardware.

The MicroSystem industry will consist of:

- customers building MicroSystems for embedded applications like automobiles, room and person monitoring, PC radio, PDAs, telephones, set top boxes, videophones, smart refrigerators.
- about a dozen foundries that fab MicroSystems.
- custom design companies that supply "core" IP and take the systems responsibility.
- existing computer system companies that have large software investments tied to particular architectures and software
- IP companies that are fab-less and chip-less that supply designs for royalty:
- ECAD companies that synthesize logic and provide design services (e.g. Cadence, Synopsis)
- circuit wizards who design: fast or low power memories (e.g. VLSI Libraries), analog for audio (also a DSP application), radio and TV tuners, radios, GPS, and microelectormechanical systems (MEMS)
- varieties of processors from traditional RISC to DSP and multimedia
- computer related applications that require much software and algorithm understanding such as communications protocols, and MPEG
- proprietary interface companies like RAMbus developing proprietary circuits and signaling standards (old style IP).

Like previous computer generations stemming from Moore's Law, a MicroSystem will most likely have a common architecture consisting of: Instruction Set Architecture (ISA) such as the 80xx, MIPS, or ARM; a physical or bus interconnect that is wholly on the chip and used to interconnect processor memory and a variety of i/o interfaces (disk, ethernet, audio,…); and software to support real time and end use applications. As in the past, common architectures are essential to support the myriad of new chips economically.

Will this new industry just be an evolution of custom microcontroller and microprocessor suppliers, or a new structure like that that created the minicomputer, PC, and workstation systems industries? Will computer companies make the transition to MicroSystems companies or will they just be IP players? Who will be the MicroSystem companies? What's the role for software companies?

---

[5] Thirty-six ECAD, computer, and semiconductor firms announced an "alliance" to facilitate building systems-on-a-chip on September 4, 1996.

## *Web Computers*

The world wide web using Internet has stimulated other computer classes to emerge, including Network Computers for corporate users, telecomputers, and television computers that are attached to phones and television sets. These near term computers use existing networks and interfaces to enhance the capability of the phone and television. In the longer term, they are integrated with all communications devices, including mobile computers and phones.

By building Web computers into telephones, TV set tops and TV sets (e.g. WebTV), and TV connected games much of the world will have instantaneous access to the web without the complexity associated with managing personal computers that will limit its use.

## *Scalable Computers Replace Non-scalable Multiprocessor Servers*

Large scale systems will be built as clusters of low cost, commodity, multiprocessor computers that communicate with one another through a fast, system area network (SAN). Clusters enable scalability to 1000s of nodes. A cluster can operate as a single system for database and on-line transaction processor (OLTP) applications. The cluster can exploit the parallelism implicit in serving multiple users in parallel or in processing large queries involving many storage devices.

Clusters will replace mainframes and minicomputer servers built as large multiprocessors with dozens of processors that share a common, high-speed bus[6]. Personal computers with only 1-4 processors are the most cost-effective nodes are dramatically less expensive, yet scalable in size, performance, location, and reliability. In 1996 (Gray, 1996), a PC cluster of several dozen nodes can perform a billion transactions per day. This is an more transaction throughput than the largest mainframe cluster.

One need for scalability comes from serving world-wide web information because web traffic and the number of users doubles annually. Future web servers will have to deliver more complex data, voice and video as subscriber expectations increase.

It's unlikely that clusters which are loosely connected computers will be a useful base for technical computing because these problems require substantial communication among the computers for each calculation. The underlying parallelism using multi-computers is a challenge that has escaped computer science and applications developers for decades, despite billions of dollars of government funding. More than likely, scientific computing will be performed on computers that evolve from the highly specialized, Cray-style, multiple, vector processor architecture with a shared memory for the foreseeable future.

## *Useful, self-maintaining computers versus users as system managers*

As the computer evolves to become a useful appliance we must remedy today's software paradox where more software provides more functions to save time, but more software increases the complexity and maintenance costs to reduce time. One of two paths may be followed, either (1) far greater complexity or (2) simplicity:

---

[6] A bus is a collection of wires used as a switch that allows processor, memory, and input-output components to communicate with one another.

1. specialized functional computers and components that know how to install and maintain themselves; This means that once a computer or a component such as a telephone, videophone, or printer arrives in an environment, such as a room, it must operate with other components reliably and harmoniously.

2. dynamically loading software from central servers to small, diskless computers such as a web terminal.

## *Telepresence* for work *is the Long-term "Killer" Application*

Telepresence is "being there, while being here at possibly some other time". Thus telepresence technology provides for both space and time shifting by allowing a user to communicate with other users via text, graphics, voice, video, and shared program operation. Communication may be synchronous with a meeting or event, or it may be asynchronous as in voice mail or electronic mail. Computers also provide for time compression since prior multimedia events can be "played back" in a non-linear fashion at rates that match the viewer's interest.

Telepresence can be for work, entertainment, education, plain communication going beyond telephony, videotelephony, mail and chat. Telepresence for work is most likely to be the "killer" app that when we look back in 2047. The question is, can mechanisms be invented for telepresence to be nearly as good as or even better than presence?

Bell characterized telepresence in four dimensions (Frankel 1996):

- mechanism: synchronous e.g. phone, chat, videophone distributed application sharing: and asynchronous such as voice mail electronic mail, video mail, web access to servers via direct use and agents. Various channels include program transfer and shared control, phone, videophone, chat windows and blackboards.

- group size & structure: 1:1 and small group meetings, 1:n presentation events

- purpose: meetings and broadcast events to interview, problem solve, sell, present, educate, operate a remote system

- work type segmented by professional discipline: engineering, finance, medicine, science, etc.

Given the modest growth in teleconferencing, and the past failures of videophones, one might be skeptical of my prediction that telepresence will be a "killer app". We are quite certain that within a decade, users will spend at least ¼ of their time, not including the time they access static web information, being telepresent. This is based on the cost of time, travel, and web access terminals coupled with the ubiquity of built in, no extra cost voice and video that can utilize POTS. In 1997 video encoding standards and products using POTS and that are compatible with telephones have just been introduced. The final telepresence inhibitor, lack of enough common platforms, explained by Metcalfe's Law will be almost entirely eliminated within a few years. Until videotelephony is ubiquitous so that everyone can communicate freely, it has little value.

## *Computers, devices, and appliances that understand speech*

In 1960, after one year of working on speech research, one of us (GB) decided to work on building computers because he predicted that it would take 20 years before any useful progress could be made. Progress has been slower than this prediction by almost a factor of two. In 1997, speech input is used for interface control and context-sensitive report generation, although speech dictation systems were available in 1990.

We believe we can optimistically assume that by 2010, speech input and output will be ubiquitous and available for every system that has electronics, including cars, computers, household appliances, radios, phones, television, toys, watches, and home or office security and environment control such as heating and lighting.

## *Video: Synthesis, Analysis, and Understanding*

The ability to synthesize realistic video in real time is the next human interface barrier. This will allow entire plays and movies to be synthetically generated. It will also allow a face-to-face Turing between a "computer synthesized image" and a person. It would seem unlikely that a computer posing as a person will be able to interact visually with a person without detection within 50 years (Kurzweil, 1990).

To illustrate evolution of a constant cost, increasing performance computer, we can look at the time when it is possible to "render and view" a movie at film resolution (approx. 20 Mpixels), in real time. Using 1994 SUN computers[7], each high-resolution film frame of Toy Story took seven hours to compute on a 165 million instruction per second (Mips) processor. Real time rendering requires a 605,000 times speed-up (7 hours/frame x 3600 seconds/hour x 24 frames/second). This requires 100 million Mips or 100 Teraops, and a computer of this speed would not be available until about 2030. However, rendering video for high definition television would require only 6 Teraops, that would be obtained 6 years earlier. Image synthesis algorithms speed up improvements are almost certain to be equal to hardware improvements so that only half this time will be required and the goal will be reached by 2010. Similarly, using special purpose rendering hardware can reduce the cost to PC price levels, provided there's a consumer desktop market e.g. games. In fact the first products using Microsoft's Talisman rendering architecture promises to generate, high resolution video of natural scenes by 1998. This will enable the desktop production of television from programs, not merely systems that store, manipulate and playback video.

## *Robots enabled by computers that see and know where they are*

The assimilation of real world data of every form, including video, global position and radar, enables new computers including useful home, office and industrial robots. Radio networks and GPS opens up more possibilities by having objects that know where they are and can report their state, that are not just adaptations of cellular phones. Everything from keys to cars and people, need not be lost.

Can useful, general purpose robots that work with everyday home or work appliances and tools such as dishwashers, brooms, vacuum cleaners, stoves, copiers, filing cabinets, construction tools and equipment, and office supply rooms be built in this short time? Or will we simply make each appliance that actually does the work more helpful? We will see a combination of the two approaches. Initially, specialized, but compatible appliances and tools will be built, followed by robots that can carry out a wide variety of activities.

## *Body Nets – Interconnecting all the computers that we carry*

A wide range of prosthetic devices are being designed, deployed and researched including artificial eyes (Dagnelie, and Massof, 1996). It is unclear when the computer will interface with humans biologically with implants to the visual cortex for artificial vision, rather than the superficial, mechanical ways they do now.

---

[7] The entire movie required 200 computers that ran two years (0.8 million hours) at a combined rate of 33 Gips.

The range of apps can vary from personal health care, control, assistance, and enhancement of human functions, to security, and communication. Wearable computers are built today to help workers operate in complex physical and logical spaces such as an airplane and wiring closet.

We can even imagine building the ultimate personal assistant consisting of "on body" computers that can record, index, and retrieve everything we've read, heard, and seen. In addition to dealing with information, the body networked monitoring computers could act as a "guardian angel."

The world-wide web offers the most potential for change at all levels of health care through standardization and universal access, including: on line information; linking human and machine created information, medical equipment, and body networked computers; caring for people by communicating with them; and on board monitoring that would warn of an event such as an impeding heart attack.

### *Computers disappear to become components for everything*

Within five years, a new MicroSystems industry will emerge. It will be based on intellectual property, that designs highly specialized nearly zero cost, systems-on-a-chip. Semiconductor foundries will build the one chip computers that have been specified by customers such as "smart appliance" manufacturers and designed by the intellectual property computer companies. These one chip, fully-networked, systems will available to be customized so that they may be used everywhere.

In 2047, the computer population is likely to be 100,000 times larger as they disappear into everything! The challenges of ubiquity through embedding into every object can positively influence computers' direction towards higher human productivity and enjoyment.

Some examples include appliances, books, pictures, and toys that communicate with one another and with us by voice, vision, and action in the context of their function. One can imagine a smart and helpful kitchen would be a dietitian, manage food (shop and control inventory), cook, serve, and clean up. If a device can be cyberized, it will.

# On predictions… and what could go wrong

Mis-predictions are legend: in 1943 Thomas Watson Sr. predicted that only five computers were needed for the country; in 1977 Ken Olsen, former CEO of Digital predicted that there would be no use for home computers. In July 1995, Bob Lucky, Vice President of Bellcore stated that: "if we couldn't predict the Web, what good are we?"

The 1962 special issue of the IRE predicted the next 50 years. Since 2010 is when semiconductor density improvement is predicted to decline or end, we can observe the progress needed to meet these early predictions.

Camras predicted a small, non-mechanical, ubiquitous memory pack, that held $10^{20}$ bits. This still appears unattainable in 2047 without some new material. He used telephony to update and communicate among the packs. He predicted home shopping, home education, and electronic payments using individual memory packs. He also predicted that consumable everyday items like food, drugs, fuel, etc. would be delivered in pipelines in suspension.

Harry F. Olson, who headed speech research at RCA's Sarnoff Labs, predicted: "There appears to be no doubt that these *(speech)* systems will be developed and commercialized because all significant steps have been made toward this goal." Three systems he described were speech to text, and speech in one language to either written or spoken speech in another language.

- Microphone > analyzer > code > typer > pages

- Microphone > analyzer > code > translator > code > typer > pages

- Microphone > analyzer > code > translator > code > synthesizer > output speech.

Simon Ramo, Founder of TRW, predicted a national network and selective databases that could be accessed by scholars, lawyers and the health care patients and servers. The simulation he prophesied for engineering design has occurred, as well as reservation and electronic payment systems.

One 1969 report for the Naval Supply Command (Bernstein, 1969), using Delphi Panel of Experts forecast the following:

- For spoken inputs, a computer will interpret simple sentences by 1975
- Some form of voice input-output will be in common use by 1978 at the latest
- Computers can be taught, thereby growing in utility by 1988
- Personal terminals that simulate activities in functional departments by 1975
- Advances in cores, wire and thin film will provide large memories with one million words by 1976
- Terabit memories at a price of 1 million dollars may be possible by 1982
- Card readers will peak at 1500 cards per minute by 1974 and then their use will decline
- Computer architecture will have parallel processing by 1975

Raj Reddy and one of us (GB) have two near term (2003) bets: AI has had as significant effect on society as the transistor, and a production model car will be available that drives itself.

Moore is unwilling to make predictions about growth beyond 2010 when various limits are reached in both materials that can resolve a bit and processing. Moore once predicted (Moore, 1980) that packaging and power supply voltages would not change from dual-in line and 5 volts.

In another case, one of us (GB) wrote about the future (Bell and McNamara, 1991) yet failed to predict the Internet. This was brought about by the serendipity of research that created a workable client-server architecture due to the standardization around the WWW, HTML, and Mosaic browser. Predictions about computer performance, structure, and applications were correct.

In predicting, the major question for 2047 is whether the technology flywheel will continue with new useful applications to sustain the investment to find more useful applications?

## Acknowledgments

The author is indebted to colleagues at Microsoft, and especially the Bay Area Research Center (BARC) researchers, and the writings and ideas of Bill Gates and Nathan Myhrvold. Sheridan Forbes of SGI helpful with the content and form. David Lyon of Cirrus Logic's PCSI division provided information about wireless networks.